# Deep learning based electrical noise removal enables high spectral optoacoustic contrast in deep tissue

Christoph Dehner,[1,2] Ivan Olefir, [1,2] Kaushik Basak Chowdhury, [1,2] Dominik Jüstel,[1,2,3,4] and Vasilis Ntziachristos[1,2,5]

[1]*Helmholtz Zentrum München, Neuherberg, Germany, Institute of Biological and Medical Imaging*
[2]*Technical University of Munich, Germany, School of Medicine, Chair of Biological Imaging*
[3]*Helmholtz Zentrum München, Neuherberg, Germany, Institute of Computational Biology*
[4]*dominik.juestel@helmholtz-muenchen.de*
[5]*bioimaging.translatum@tum.de*

**Abstract:** Image contrast in multispectral optoacoustic tomography (MSOT) can be severely reduced by electrical noise and interference in the acquired optoacoustic signals. Signal processing techniques have proven insufficient to remove the effects of electrical noise because they typically rely on simplified models and fail to capture complex characteristics of signal and noise. Moreover, they often involve time-consuming processing steps that are unsuited for real-time imaging applications. In this work, we develop and demonstrate a discriminative deep learning (DL) approach to separate electrical noise from optoacoustic signals prior to image reconstruction. The proposed DL algorithm is based on two key features. First, it learns spatiotemporal correlations in both noise and signal by using the entire optoacoustic sinogram as input. Second, it employs training based on a large dataset of experimentally acquired pure noise and synthetic optoacoustic signals. We validated the ability of the trained model to accurately remove electrical noise on synthetic data and on optoacoustic images of a phantom and the human breast. We demonstrate significant enhancements of morphological and spectral optoacoustic images reaching 19% higher blood vessel contrast and localized spectral contrast at depths of more than 2 cm for images acquired in vivo. We discuss how the proposed denoising framework is applicable to clinical multispectral optoacoustic tomography and suitable for real-time operation.

## 1. Introduction

Electrical noise is a key source of signal corruption in optoacoustic imaging and arises due to thermal effects (thermal noise) and from electromagnetic interference (parasitic noise); the latter possibly generated by the optoacoustic system itself or the environment [1]. While thermal noise can be modeled as white Gaussian noise [2], parasitic noise entails complex spatiotemporal correlations, and thus cannot be efficiently captured by an analytical model [1]. Both thermal and parasitic noise cause artifacts in reconstructed optoacoustic images that severely decrease morphological and spectral contrast. Whereas shielding hardware can suppress some parasitic noise, this solution is device-specific and often incomplete [1]. Signal processing techniques, which are applicable across platforms, are therefore needed to remove both parasitic and thermal noise.

Noise in optoacoustic images hinders the detection and identification of fine structures in tissue, particularly as the signal-to-noise ratio (SNR) in the data decreases with increasing depth [3]. Besides the reduction of image contrast, noise also challenges the quantification and spectral un-mixing of optoacoustic images acquired at multiple wavelengths [4, 5]. In particular, corrupted spectral information decreases the spatial resolution of multispectral optoacoustic tomography (MSOT) because it necessitates averaging the spectra obtained from large tissue regions for reliable quantification [6-8]. Efficient noise reduction algorithms are therefore critical for improving the performance of spectral optoacoustics.

Frequency filtering using band-pass filters cannot adequately separate thermal and parasitic noise from optoacoustic signals because the frequency content of optoacoustic signals and noise overlap significantly. For this reason, data averaging and regularization methods have been more commonly considered to minimize the effects of electrical noise from optoacoustic tomographic images [3, 7-11]. While data averaging effectively reduces zero-mean electrical noise, combining multiple acquisitions reduces imaging rates and increases vulnerability to motion artifacts, particularly in clinical applications or when using a handheld system. Regularization of model-based reconstructions may decrease the effects of electrical noise, but this reduction is either limited to the noise characteristics captured by the regularization functional or realized at the cost of data fidelity, thereby corrupting the meaningfulness of the reconstructed image. Furthermore, iterative model-based reconstruction is computationally intensive, and therefore unsuitable for applications that require real-time feedback [4, 12, 13]. Another approach to reduce noise is based on sparse (typically Wavelet-based) representations for optoacoustic signals [1, 14, 15]. Noise is assumed to distort the sparsity properties of optoacoustic signals, which enables its removal, e.g., via thresholding techniques. However, the denoising performance of such methods is limited by their reliance on oversimplified models of noise and optoacoustic signals.

Recently, discriminative deep neural network models have achieved state-of-the-art performance on general image denoising tasks, like Gaussian denoising, deblocking, super-resolution, inpainting, and dehazing [16-19]. These deep neural network models capture the required data transformations for denoising in a data-driven way by utilizing large ground truth training datasets. In this way, discriminative deep neural networks are capable of more accurate, robust, and fast denoising than traditional methods that rely on rigid analytical models because they can adjust to complex data characteristics during training and are evaluated in real-time on modern GPUs.

In this work, we examine whether discriminative deep learning can separate thermal and parasitic noise from optoacoustic signals. We show that the modelling capabilities and the computational efficiency of a deep neural network facilitates denoising in optoacoustic tomography that is both precise enough to remove noise with complex characteristics and fast enough for real-time imaging applications. We design a deep neural network model to simultaneously denoise the entire sinogram of an optoacoustic scan, i.e. the complete dataset acquired from all transducers. Entire sinogram denoising enables the network to capture spatiotemporal correlations within both parasitic noise and true optoacoustic responses, and thus more efficiently separate the two. Exploiting the independence of electrical noise and optoacoustic signals, we train the network on a large ground truth dataset of experimentally acquired pure noise and synthetic optoacoustic sinograms. We validate that the model removes thermal and parasitic noise from both synthetic sinograms and optoacoustic images of a phantom. We lastly apply the trained model to clinical MSOT images of breast tissue and show significant enhancements in morphological and spectral contrast. The improved contrast allows for tissue components to be more accurately localized and quantified and yields more meaningful correlations with the spectra of know absorbers in tissue, thereby increasing access to endogenous biomarkers in deep tissue, such as breast vasculature or hemoglobin contrast inside a cancerous tumor.

## 2. Methods

In the following, we formalize our methodology for removing electrical noise from optoacoustic sinograms. First, we reformulate the denoising problem as a decomposition task.

Based on this formulation, we derive a discriminative deep learning framework for denoising optoacoustic sinograms. At the end of the chapter, we explain the experiments that we use to validate this approach.

### 2.1 Denoising via decomposition

An optoacoustic scan at a fixed excitation wavelength consists of measured optoacoustic pressure signals $s_d[t]$, indexed by time samples $t \in [1, 2, \dots, N_\text{samples}]$ and transducer locations $d \in [1, 2, \dots, N_\text{transducers}]$. The collection of signals recorded at all the transducers compose the sinogram $s[d, t] \coloneqq s_d[t]$ of the scan. We model the measured optoacoustic sinograms probabilistically as samples $s$ of a random field $S$. The main assumption that leads to the formulation of denoising as a decomposition problem is that $S$ is a sum of two independent random fields $S_{OA}$ and $S_{noise}$, which describe the signal content due to optoacoustic responses and electrical noise, respectively. This assumption is justified by the fact that electrical noise in optoacoustic tomography typically originates from common system thermal noise and electromagnetic inference that is not influenced by the optoacoustic signal [1]. The probability distributions underlying $S_{OA}$ and $S_{noise}$ are denoted by $P_{OA}$ and $P_{noise}$:

$$S = S_{OA} + S_{noise}$$
with $S_{OA} \sim P_{OA}$ and $S_{noise} \sim P_{noise}$ independent. (1)

Optoacoustic scans at different wavelengths are modelled as independent realizations of $S$ because the noise is caused by the electronics of the imaging system that are not affected by the wavelength switching of the laser. In summary, isolating the noise-free optoacoustic sinogram $s_{OA}$ given $s$ is equivalent to decomposing $s = s_{OA} + s_{noise}$ into its two components $s_{OA}$ and $s_{noise}$.

### 2.2 Deep learning based denoising framework

To solve this decomposition problem, we need access to the distributions $P_{OA}$ and $P_{noise}$. However, both random fields $S_{OA}$ and $S_{noise}$ are non-homogeneous and anisotropic with intricate spatial and temporal correlations due to the physics underlying the signals and the fact that electrical noise in optoacoustic systems often contains complex parasitic noise (Fig. 2c) [1]. Even if accurate models for $P_{OA}$ and $P_{noise}$ could be explicitly formulated, the high-dimensionality of the problem would make them difficult to handle. We therefore present a data-driven approach that allows us to rely on the empirical distributions of $P_{OA}$ and $P_{noise}$ via sampling of $S_{OA}$ and $S_{noise}$. We first explain sample acquisition and then elaborate on our methodology for solving the decomposition task.

Because of the independence of $S_{OA}$ and $S_{noise}$ and the wavelength independence of $S_{noise}$, electrical noise can be directly measured in the absence of any absorbers that would emit optoacoustic responses. We thus obtained samples of the electrical noise distribution $S_{noise}$ of the test system by immersing the scanner in a water tank and measuring from 700 to 790 nm, where light absorption in water is negligible.

Acquiring samples of $S_{OA}$ in an experimental setup is a laborious and time-consuming process that requires averaging multiple scans of the same location to remove electrical noise. Additionally, patient or operator movement impedes accurate averaging. Therefore, instead of experimentally acquiring noise-free optoacoustic sinograms, we generated samples of $P_{OA}$ via simulation by applying an accurate acoustic forward model of the scanner [20, 21] to publicly available images from the PASCAL VOC2012 dataset [22], a diverse collection of over 17 000 images covering a large range of features. Utilizing these images as underlying initial pressure distributions in the simulations enables us to account for a broad range of potential features in optoacoustic sinograms and should yield a good approximation of the empirical distribution of $P_{OA}$. In addition, the general scope of the training data ensures that the denoising model is universally and with uniform performance applicable to arbitrary scans acquired by the considered system.

Using the samples from $S_{OA}$ and $S_{noise}$, we utilize a Unet-like deep neural network [23] to solve the decomposition task at hand. Fig.1 depicts the deep learning based approach. Fig.1a shows the training setup. We iteratively train the network on randomly selected pairs of samples $s_{OA}$ and $s_{noise}$ from $S_{OA}$ and $S_{noise}$ to infer the noise component ($s_{noise}$) from a noisy input sinogram $s = s_{OA} + s_{noise}$. In this way, the network is optimized to adopt the complex characteristics of $P_{OA}$ and $P_{noise}$. Fig. 1b depicts the evaluation setup. Once trained, we can use the neural network to infer electrical noise from noisy input scans. A detailed description of the deep neural network and all used training parameters are included in the supplemental information.

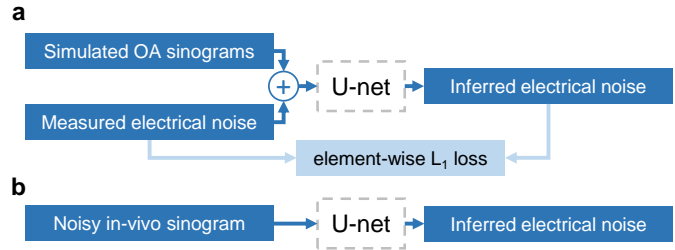

Fig. 1. Discriminative deep learning framework for denoising optoacoustic sinograms. a) Training setup of the method. The network was trained iteratively using simulated samples from the optoacoustic sinogram distribution of the test system and experimentally acquired samples of the electrical noise distribution of the system. b) Evaluation setup of the method. The trained neural network can infer the electrical noise from a noisy input sinogram. Subtracting the inferred noise from the input sinogram yields the denoised sinogram.

## 2.3 Experiments

As a test system for the denoising algorithm, we used a custom prototype of an MSOT Acuity Echo handheld scanner (iThera Medical GmbH, Munich, Germany; see supplementary information for detailed specifications). Fig. 2 provides an overview of the imaging system and its output. Fig. 2a shows the scanning procedure using the handheld imaging probe of the system. Fig. 2b illustrates the data layout of a multispectral stack acquired by the imaging system. A multispectral stack consists of 28 sinograms recorded at wavelengths from 700 – 970 nm in steps of 10 nm. Fig. 2c shows electrical noise from an exemplary optoacoustic in vivo scan. We observed that electrical noise in the system consists of two additive components: normally distributed thermal noise with a mean of zero and a standard deviation in the range of 0.2 – 0.3 and complex parasitic noise that is presumably caused by the switching-mode power supply of the system (examples marked with red arrows in Fig. 2c).

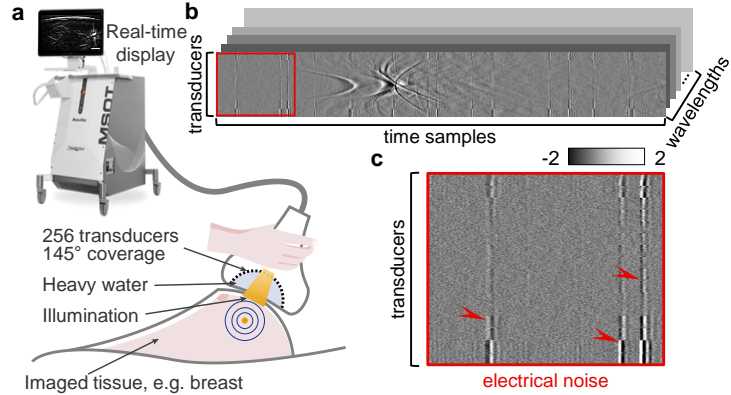

Fig. 2. Overview of the handheld MSOT system and its output used to evaluate the proposed denoising method. a) Illustration of the scanning procedure using the handheld imaging probe of the test system. b) Data layout of a measured multispectral stack of sinograms. The depicted sinogram shows the recorded signals during an exemplary scan of a human breast lesion at 960 nm. c) Magnification of the marked signals in b, which were recorded prior to responses from tissue and thus are predominately comprised of electrical noise.

We first evaluated the ability of the proposed deep learning framework to remove the combination of Gaussian thermal and complex parasitic electrical noise observed in the test system. We trained and evaluated a deep neural network on experimentally acquired samples of the electrical noise distribution $S_{noise}$ of the system and simulated samples of the optoacoustic signal distribution $S_{OA}$ (denoted as Dataset-EN). Next, we applied the trained denoising network to measurements of a phantom (denoted as Dataset-Ph) comprising a copper sheet and tubes filled with ink that are embedded in agar layers with slightly different speeds of sound (Fig. 4a, inset, see supplementary information for detailed description). To evaluate the denoising performance of the framework on in vivo scans, we subsequently applied the trained deep neural network to 81 multispectral optoacoustic scans of human breast cancer lesions (denoted as Dataset-BC). These scans were acquired in a study that was approved by the local ethics committee of the Technical University of Munich (27/18 S). All participants gave written informed consent upon recruitment.

In a second experiment, we sought to evaluate the applicability of the method to optoacoustic signals that do not suffer from parasitic noise with spatiotemporal correlations at all (i.e. only from white Gaussian thermal noise) by training an additional deep neural network model on simulated samples from $S_{OA}$ that were corrupted by different levels of white Gaussian noise (denoted as Dataset-GN). A summary of the four datasets as well as all applied pre-processing steps are given in the supplementary information.

### 3. Results

The proposed deep learning framework for denoising optoacoustic sinograms significantly improves the SNR of both simulated and in vivo data in real-time. Based on the resulting increased quality of the optoacoustic signal data, the denoising method enables improved optoacoustic image contrast and spectral unmixing performance. In the following, we report the detailed findings in the signal, image, and spectral domains.

*3.1 Denoising performance in the signal domain*

Optoacoustic signals and electrical noise are both complex broadband signals whose characteristics overlap significantly. The presented data-driven approach can disentangle these overlaps by accessing and separating the data manifolds of signal and noise in sinograms. We observed significant reductions in noise levels, both visually and quantitatively, upon application of the denoising method to sinograms that were corrupted by a combination of Gaussian thermal and complex parasitic electrical noise (Dataset-EN and Dataset-BC), as well as to sinograms that were corrupted only by Gaussian noise (Dataset-GN). Fig. 3 summarizes these results. We begin by visually inspecting the effects of the denoising method for an exemplary scan of a breast lesion in Fig. 3a-f. Fig. 3a shows the noisy sinogram before denoising. Because of the radial nature of wave propagation and the circular shape of the used imaging probe, optoacoustic responses appear as bow-shaped structures in the sinogram. The sinogram is distorted by additive electrical noise that is composed of zero-mean Gaussian noise and complex noise artefacts with strong spatiotemporal correlations (as also shown in Fig. 2c). Fig. 3b depicts the electrical noise component inferred by the trained neural network, which demonstrates the network's ability to extract electrical noise. Finally, Fig. 3c shows the denoised sinogram, which was obtain by subtracting the inferred noise from the recorded sinogram. Figs. 3d-f depict enlargements of identical temporal sections of the images in Figs. 3a-c, highlighting the fine features of the optoacoustic signal that are exposed upon removal of electrical noise (yellow arrows).

Fig. 3g-i provide an in-depth quantitative analysis of the of the network's denoising performance. These results confirmed the ability of the network to consistently remove electrical noise with high accuracy from both synthetic and in vivo optoacoustic sinograms. Fig. 3g compares the distributions of SNRs within the test split of Dataset-EN before and after denoising (see supplementary information for a detailed definition of the SNR of sinograms). Application of the denoising method to the sinograms in Dataset-EN resulted in an average improvement in SNR of 10.9 dB, with improvements for individual sinograms ranging from 4.6 dB to 20.0 dB. After the neural network was trained and tested on Dataset-EN, we applied it to denoise scans of breast lesions (Dataset-BC) to demonstrate its applicability to in vivo data. Fig. 3h shows a plot of the mean SNRs ($SNR_{mean}$, see supplementary information for the definition) of these individual time samples from Dataset-BC before and after application of the network, which indicates a time-independent increase in $SNR_{mean}$ of approximately 20.8 dB after denoising. The uniformity of the increase in $SNR_{mean}$ demonstrates that the trained neural network can extract electrical noise both from strong optoacoustic responses in superficial tissue (time samples 400-700), as well as from signals deeper in tissue, which have lower amplitudes due to light fluence attenuation (time samples 1100-1400).

Furthermore, we demonstrated the ability of the method to compensate for the variations in parasitic electrical noise within the transducer array of the test system (see Fig. 2c for details) to confirm the applicability of the trained network to in vivo scans. For that, we calculated the $SNR_{mean}$ for Dataset-BC individually for all transducer elements, rather than for the whole sinograms, before and after denoising. As shown in Fig. 3i, applying the trained network to the breast scans from Dataset-BC improves the $SNR_{mean}$ at all transducers by an average of 22.4 dB in a near uniform manner. Note that the transducers at the boundaries of the detector probe have lower SNRs than the central transducers due to the probe layout partially shielding the outermost transducers from arriving acoustic waves. The lower $SNR_{mean}$ values at transducers 30-43, 79-87, 167-175, 213-227 result from acoustic noise waves propagating along the transducer array, which corrupts the ground truth noise estimation used to calculate the $SNR_{mean}$ (see Equation 1 in the supplemental information). These noise waves depend on the imaged tissue, and therefore cannot be removed by the neural network.

Thus far, we have demonstrated the ability of the denoising method to accurately isolate the electrical noise of the test system composed of Gaussian thermal and complex parasitic noise. Since electrical noise is hardware-dependent and thus varies between different imaging systems, we utilized Dataset-GN to test the generalizability of the denoising framework to other imaging systems with a different amount of Gaussian thermal noise and a potentially lower amount of parasitic noise (i.e. no parasitic noise at all). Note that the standard deviations of the white Gaussian noise in the test split of Dataset-GN were four times as high as in the train split, therefore challenging the robustness of the trained neural network to generalize to increased amounts of Gaussian noise, e.g. caused by changes in hardware configurations or environmental conditions such as humidity and temperature. Fig. 3j summarizes the denoising performance of the trained network applied to Dataset-GN, which afforded improvements in SNR for all tested standard deviations of white Gaussian noise. These results demonstrate that the denoising approach can accurately isolate electrical noise that does not comprise parasitic noise with strong spatiotemporal correlations.

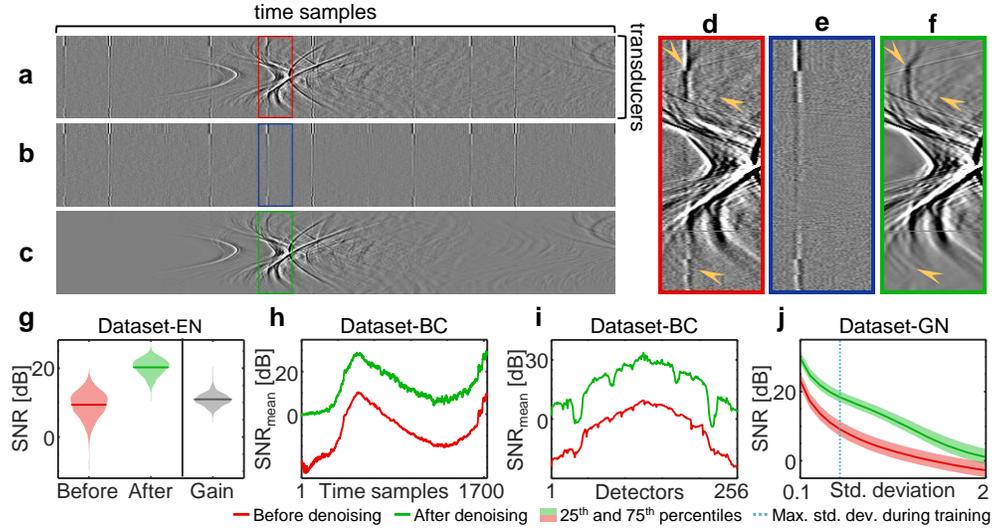

Fig. 3. Evaluation of the proposed denoising approach in the signal domain. a) Noisy sinogram from an exemplary scan of a human breast lesion. b) Electrical noise component inferred by the neural network. c) Denoised sinogram obtained by subtracting b from a. d-f) Magnifications of the marked areas in a-c. g-j) Quantitative evaluation of the denoising performance. g) Comparison of the SNR distributions in simulated optoacoustic sinograms that are distorted by electrical noise before and after denoising. The mean gain is 10.9 dB. h-i) Evaluation of *in vivo* scans of human breast lesions. h) Mean SNR ($SNR_{mean}$) of individual time samples. The average increase is 20.8 dB. i) Individual $SNR_{mean}$ of all detectors. The average increase is 22.4 dB. j) Average SNRs before and after denoising of simulated optoacoustic sinograms that are distorted by different levels of zero-mean Gaussian noise. During training, standard deviations up to 0.5 were used. One of the detectors (no. 61) was defective and excluded from the plots in g-j.

## 3.2 Denoising enables high contrast in optoacoustic images

Thus far, we have demonstrated the ability of the denoising network to isolate and remove electrical noise in optoacoustic sinograms. In this section, we analyze the effects of denoising on reconstructed optoacoustic images. First, we ensure that the denoising network successfully removes noise artifacts without distorting any true optoacoustic image structures using optoacoustic images of a phantom. Subsequently, we evaluate the improved image contrast due to denoising in a clinical dataset of scans of human breast lesions, to show the potential for improved diagnostic capability of optoacoustic tomography.

We utilized a model-based inversion algorithm (see supplementary information for details) to reconstruct optoacoustic images (i.e. initial spatial pressure distributions) from all scans in the datasets Dataset-Ph and Dataset-BC, both with and without denoising the recorded sinograms with the trained neural network. Fig. 4a-e illustrates the qualitative improvements to selected images upon application of the neural network. Fig. 4a shows an optoacoustic image of a phantom at 700 nm, reconstructed from a noisy sinogram. Zero-mean Gaussian noise in the recorded sinogram reduces the overall contrast in the image, whereas parasitic noise leads to ring artifacts that obscure potentially relevant image features. The arrangement of the phantom is shown in the inlay of Fig. 4a. Fig. 4b depicts the same optoacoustic image reconstructed from a denoised sinogram, demonstrating that the neural network can significantly reduce both the background noise and the ring artefacts. Fig. 4c plots the difference between Figs. 4a and b, which yields artifacts and background noise but no real structures, emphasizing the ability of the network to accurately identify and isolate noise in optoacoustic images. Fig. 4d and e show the optoacoustic images of a malignant breast tumor at 870 nm. The denoised image in Fig. 4e appears significantly richer in contrast than the original image in Fig. 4d and contains structures that are not visible prior to the denoising. To highlight the clinical relevance of the improved contrast, note that in Fig. 4e, the optoacoustic contrast inside the tumor core regions (outlined in blue) is separated from the noise that dominates these regions in Fig. 4d.

Next, we evaluated the contrast resolution (CR) of blood vessels in the optoacoustic images reconstructed from Dataset-BC to quantify the enhancement capabilities of the trained neural network in the image domain. Blood vessels and background ROIs were first manually segmented, as depicted in Fig. 4f, and used to calculate the blood contrast resolution (see supplementary information for details). The distributions of contrast resolution in Dataset-BC before and after denoising are compared in Fig. 4g, which shows an average improvement of 0.083 with a range of 0.003 to 0.55 for individual images. As shown in Fig. 4h, the average improvement in blood contrast resolution is consistent across all wavelengths, demonstrating the network's ability to remove noise, independent of the varying strength of individual absorbers across the accessible spectrum.

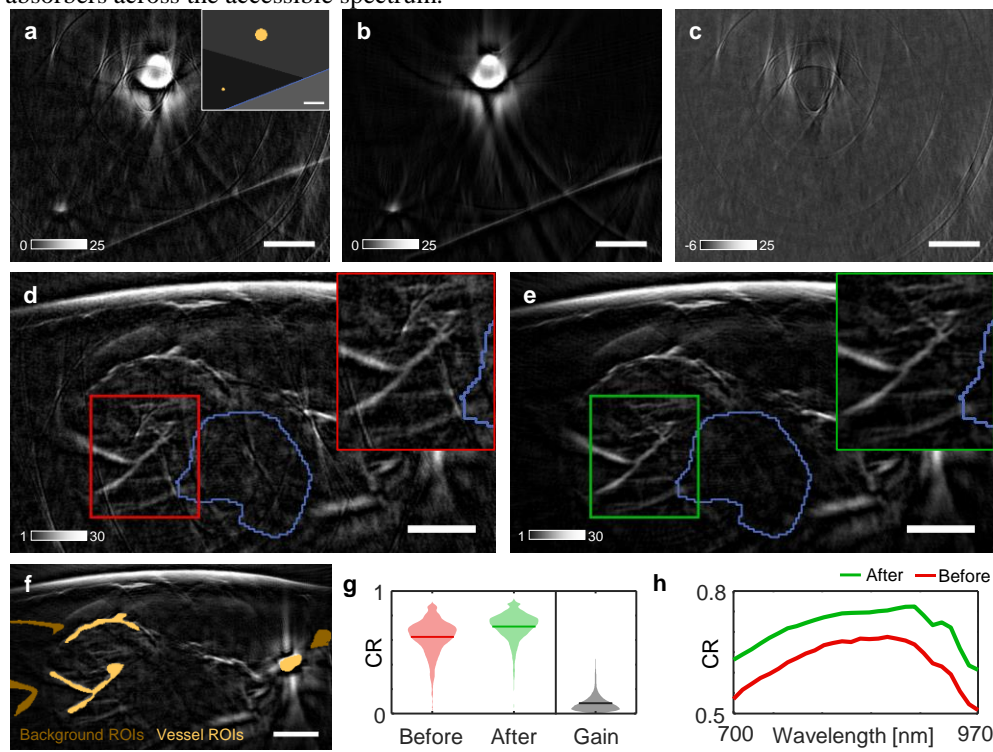

Fig. 4. Demonstration of improved image quality in denoised scans of a phantom and of human breast lesions. a) Optoacoustic image of a phantom before denoising. The overlayed image shows the arrangements of the phantom: tubes filled with ink (yellow), copper sheet (blue), and agar layers with slightly different speed of sound distributions (grey). b) Corresponding optoacoustic image after denoising. c) Difference of a and b. d-e) Optoacoustic image of a malignant breast tumor d) before and e) after denoising. The location of the hypoechoic tumor

core, obtained from ultrasound images, is outlined in blue. f) Examples for the vessel and background ROIs that are used to compute the contrast resolution. g-h) Quantification of the contrast resolution (CR) of blood vessels in scans of breast lesions before and after denoising. The average increase is 0.083. The minimal gain is 0.003. The depicted optoacoustic images of the phantom and the breast lesion are obtained at 700 nm and 870 nm, respectively. All scale bars correspond to 5 mm.

### 3.3 Deep learning based denoising enables high spectral contrast

A further central finding of this work is the ability of the presented denoising approach to significantly improve spectral contrast in MSOT, i.e. the differentiation of chromophores based on their absorption spectra. We found that upon application of the denoising method to the MSOT scans from Dataset-BC, the dominant absorbers in breast tissue – hemoglobin, lipids, and water – are more accurately identified and localized. To analyze the spectral contrast, we applied blind spectral unmixing via non-negative matrix factorization (NMF, see supplementary information for details) to the original and denoised breast images and decomposed each of the two datasets into 10 spectra and corresponding unmixing coefficients. Note that unlike linear unmixing based on the reference absorption spectra of chromophores in tissue, NMF finds both the spectra and unmixing coefficients in a data driven way and thus extracts variants of the reference spectra that consider effects from spectral coloring.

Fig. 5 compares the spectral contrast of the original and denoised MSOT breast images from Dataset-BC. In Fig. 5a-c, we show the NMF spectra obtained from the original (Fig. 5a) and from the denoised (Fig. 5b) data next to the reference absorption spectra of the most prominent chromophores in tissue (oxygenated and deoxygenated hemoglobin, water, and lipids, see Fig. 5c). The spectra derived from the original data show a significant number of sharp peaks attributable to ring artifacts from parasitic noise, rather than specific absorbers in tissue. In contrast, the spectra obtained from the denoised images are broader, smoother, and are more easily related to the reference absorption spectra of hemoglobin (spectra no. 1, 2, 6, 7, 8, 10), fat (spectra no. 4, 9), and water (spectra no. 3, 5). The increased number of meaningful spectra found by NMF demonstrates superior spectral contrast of the denoised images compared to the original images.

In Fig. 5d-g, we visualize the obtained spectral decompositions before and after denoising of an exemplary multispectral stack to visually confirm the ability of the network to enable better spectral contrast. We color-encode and blend the unmixing coefficients of three NMF spectra at a time, which correlate with reference absorption spectra of hemoglobin (Fig. 5d,e) and of lipids and water (Fig. 5f,g), covering approximately the same spectral regions for the original and the denoised data. To improve the dynamic range of the rendered images, we display the square roots of all coefficients in the visualizations. Whereas the visualizations derived from the original data are dominated by overlapping coefficients of different spectra (appearing as white in the color-encoding) and by ring noise artifacts (example marked with white arrows in Fig. 5g), the visualizations derived from denoised data show a reduction of noise artifacts and express significantly richer rich spectral contrast. In addition, while the tumor core (outlined in white) in Fig. 5d,f contains a lot of noise, this noise is removed by the denoising method in Fig. 5e,g, revealing hemoglobin contrast inside the tumor (white arrows in Fig. 5e). In summary, improved spectral contrast is observable in two ways upon application of the denoising method to the scans from Dataset-BC: First, blind spectral unmixing retrieves a more versatile set of spectral components and second, the denoising method enables a more meaningful decomposition of the acquired images into the found spectra.

Finally, we evaluated the effects of denoising on the spectral contrast of deep tissue by comparing how spectral coloring, i.e. depth dependent variation of observed spectra due to fluence attenuation, is captured in the obtained spectral decompositions of the original and denoised multispectral stacks from Dataset-BC. For that, we selected all NMF spectra that correlated to the absorption spectra of hemoglobin. We then plotted the individual relative contributions of those spectra to the hemoglobin contrast as a function of depth for both the original and denoised breast images (Fig. 5h-i, see supplementary information for detailed explanation). The relative contributions of the three spectra that comprise hemoglobin contrast in the original images (Fig. 5h, spectra no. 1, 2, 6 from Fig. 5a) are almost constant after depths greater than 1.5 cm (corresponding to approximately 0.5 cm of the coupling medium and 1 cm of tissue). In contrast, the relative contributions of the six NMF spectra that comprise blood contrast in the denoised data (Fig. 5i, spectra no. 1, 2, 6, 7, 8, 10 from Fig. 5b) vary with depth up to 3.5 cm (corresponding to approximately 0.5 cm of the coupling medium and 3 cm into the tissue) in a manner attributable to spectral coloring. The ability to identify these variations upon application of the denoising network demonstrates that the presented method enhances the spectral contrast of MSOT at all depth levels and thus also enables more accurate imaging of deep tissue.

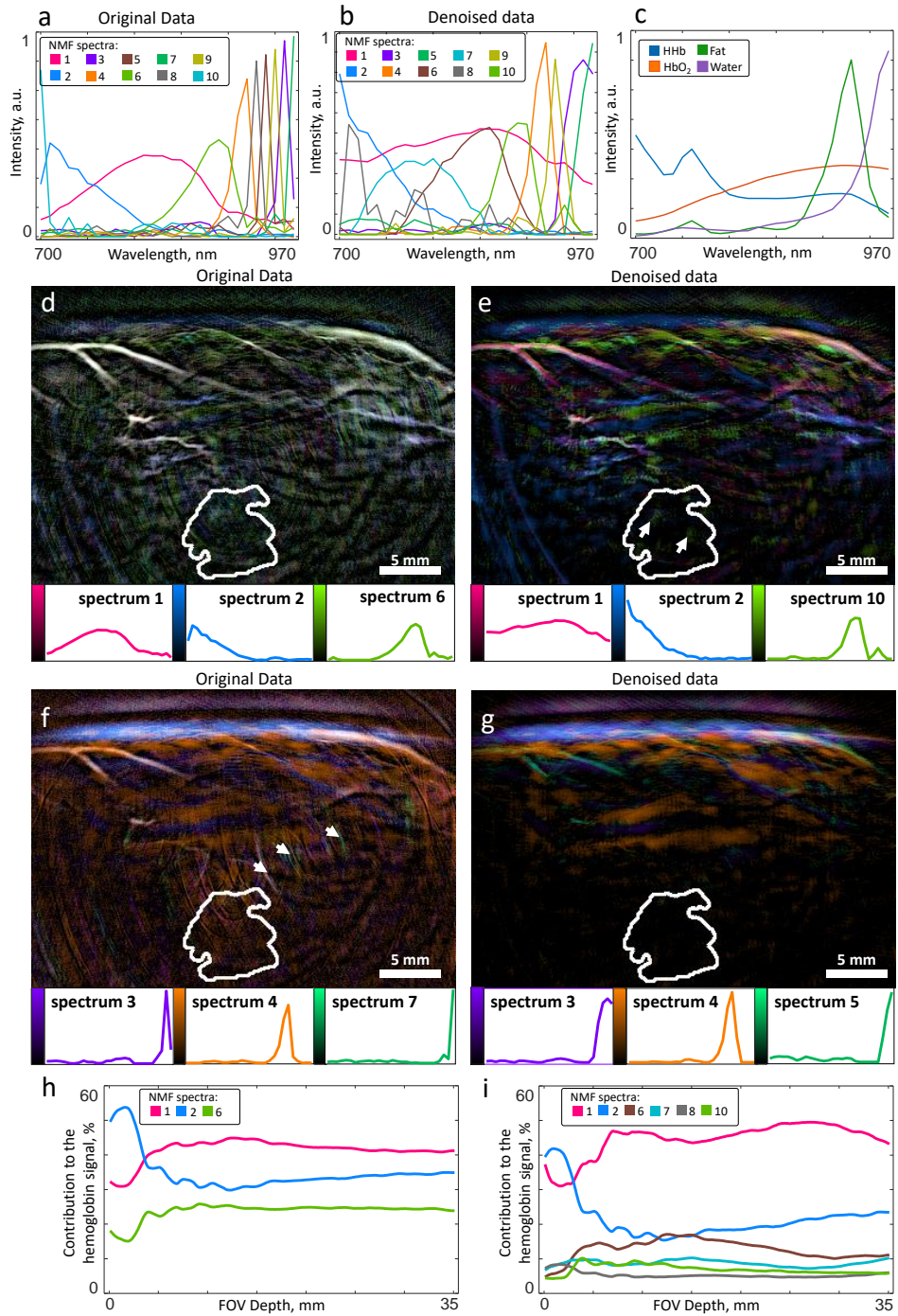

Fig. 5. Effects of denoising on the spectral content of optoacoustic images. a-b) NMF spectra that were obtained from the original (a) and the denoised (b) MSOT images of human breast lesions from Dataset-BC. c) Reference absorption spectra of the most prominent chromophores in breast tissue. d-g) Visualizations of the NMF decomposition of an exemplary MSOT image showing a malignant breast tumor at approximately 2 cm depth before (d, f) and after (e, g) denoising. The images color-encode the contributions of respectively three spectra that correlate with the absorption spectra of hemoglobin (d, e) and with fat and water (f, g). The position of the tumor, obtained from ultrasound images, is demarcated by the white outlines. h-i) Relative contributions of the NMF spectra that correlate to hemoglobin absorption spectra to overall hemoglobin contrast in the breast images from Dataset-BC before (h) and after (i) denoising as a function of depth.

## 4. Discussion

Optoacoustic signals are relatively weak and thus susceptible to corruption by electrical noise during the imaging process, which impedes morphological and spectral contrast. In this work, we presented a discriminative deep learning based denoising method for optoacoustic sinograms, which employs a deep neural network trained on samples of experimentally acquired electrical noise and simulated ground truth optoacoustic signals. We demonstrated that the trained deep neural network could accurately remove electrical noise from in vivo scans of a MSOT system. The proposed signal processing technique offers a fast and accurate approach to improve the SNR of recorded optoacoustic sinograms, increase morphological image contrast, and enable rich spectral contrast at high resolution in handheld MSOT imaging.

The presented deep learning based denoising framework is effective because it can access the topology and the statistics of pure electrical noise and optoacoustic signal datasets. This structural information contained in large datasets has recently been made accessible by advances in computational power and methodology and is the driving force behind the increasing success of deep learning methods in medical imaging [24, 25]. We generated such a large and high-quality dataset by complementing the experimentally acquired pure noise data with simulated optoacoustic sinograms. The simulated data was obtained by applying a mathematical model of the imaging system to a general feature image database, thereby incorporating prior knowledge about the imaging system without sacrificing general applicability of the method to any data acquired by the system. This is an example of the integration of a physical model into data-driven methods, which remains a major challenge in machine learning [26-28]. The proposed method allows a tradeoff between model accuracy and generality. For example, one could potentially enhance denoising performance by selecting an optoacoustic signal dataset that more specifically reflects typical tissue responses. However, the method achieves accurate denoising and good generalization beyond the training data without any such specialization.

Furthermore, the trained deep neural network model provides a means of fast denoising. Clinical optoacoustic imaging systems typically provide real-time feedback to the device operator on a built-in monitor. Due to the restricted processing times, these online images are usually much lower in quality than those produced offline, which can lead to longer imaging sessions and incorrect selection of regions of interest. We demonstrated that the method can denoise a full optoacoustic sinogram of the MSOT system in approximately 9 milliseconds, which is fast enough for real-time feedback during device operation. Improving instantaneous image quality enhances the dynamic imaging capabilities of MSOT [29] while decreasing examination times.

In addition to better image quality, the denoising method also enhances the fidelity of the obtained spectral information. MSOT enables molecular contrast by extending the high-resolution optical contrast of optoacoustic imaging to the spectral dimension [30]. However, previous clinical MSOT studies extracted spectral information mostly by averaging over larger areas in MSOT images [6-8], thereby sacrificing the superior resolution of optoacoustics. In this work, we demonstrated that denoising overcomes the necessity to average over large tissue regions and enables spectral contrast down to the system resolution, which is ~200 μm in the test system. High-resolution spectral contrast was highlighted by localizing hemoglobin contrast inside a 2 cm deep breast tumor. Spectral contrast is of the utmost interest for clinical applications of MSOT, since it, for example, enables in detail studies of local blood oxygenation and tissue metabolism.

Finally, the presented denoising framework is also applicable to other (optoacoustic) imaging systems. For example, optoacoustic mesoscopy [31] and microscopy systems [32] are beset with similar electrical noise, making the approach of acquiring pure noise measurements and simulating signals with a numerical model applicable to these systems without any major changes. Other noise sources, like speckle noise in ultrasound imaging [33] or optical coherence tomography [34] and shot noise in coherent diffraction imaging [35] can be modelled as independent multiplicative noise and can thus be approached by adapting the proposed method accordingly. More generally, the presented methodology can in any context disentangle two independent random fields that are mixed in a known way and whose distributions can be accessed by sampling.

In summary, the deep learning framework that we propose in this work is an efficient and flexible method for denoising optoacoustic tomography data. By significantly improving the data quality of the considered MSOT system, we moved one step closer to the full potential of handheld MSOT imaging, which is dynamic high-resolution molecular contrast deep in tissue.

**Funding.** This project has received funding from the European Research Council (ERC) under the European Union's Horizon 2020 research and innovation programme under grant agreement No 694968 (PREMSOT), from the European Union's Horizon 2020 research and innovation programme under grant agreement No 862811 (RSENSE), and by the Deutsche Forschungsgemeinschaft (DFG), Sonderforschungsbereich-824 (SFB-824), subproject A1.

**Acknowledgments.** The authors would like to thank Hong Yang for providing and scanning the phantom, Angelos Karlas and Korbinian Paul-Yuan for designing the study to acquire the breast scans, Stefan Metz for acquiring the breast scans, Jan Kukačka for his help with the analysis of the MOST breast images, and Dr. Robert Wilson for his attentive reading and improvements of the manuscript.

**Disclosures.** Vasilis Ntziachristos is an equity owner and consultant for iThera Medical GmbH.

**Data availability.** Data underlying the results presented in this paper are not publicly available at this time but may be obtained from the authors upon reasonable request.

## 5. Supplemental material

### 5.1 Neural network design and training

A detailed illustration of the proposed neural network is given in Supplementary Fig. 1. We adopted the U-Net neural network architecture [23], which has been successfully used to solve decomposition tasks in other medical applications [36]. The U-Net architecture incorporates key concepts of state-of-the-art methods for general image denoising [17, 18], such as an encoder-decoder based architecture, batch normalization and skip connections. Analogous to the DnCNN denoising model [17], we apply residual learning [37] to minimize the necessary expressiveness of the network. We therefore designed the network to infer only the electrical noise $s_{noise}$ from a noisy input sinogram $s$. Based on insights from [38], we utilize the L1 norm of the difference of inferred and ground truth noise as loss functional. We iteratively train the network for 300 epochs using the ADAM optimizer [39] with batch size $= 1$ and momentum parameters $\beta_1 = 0.5$ and $\beta_2 = 0.999$. The learning rate is set to $0.0001$ and is linearly decreased to zero in the last 50 epochs. To accelerate the learning process, we scale all input data of the neural network by a constant factor of $0.004$ to achieve a signal range of $[-1; 1]$. After having passed the neural network, all signals are rescaled to the original range. In total, one training process takes approximately three days on an NVIDIA GeForce RTX2070 GPU. After training, we select the final model based on the minimal loss on a validation split of the dataset.

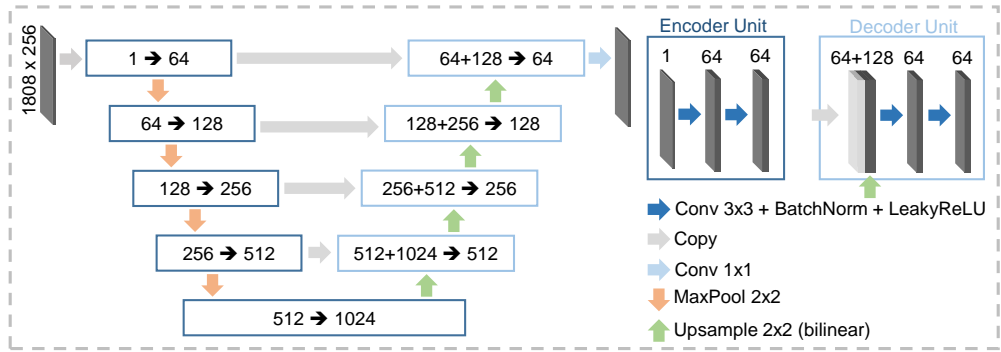

Fig. S1. Encoder-decoder based U-net architecture of the deep neural network.

### 5.2 Handheld MSOT test system

As test system for the denoising algorithm, we used a custom prototype of an MSOT Acuity Echo handheld scanner (iThera Medical GmbH, Munich, Germany). The system is equipped with a tunable laser that illuminates tissue with laser pulses of ~8ns duration with an energy of 16mJ and a repetition rate of 25 Hz, thereby staying within the energy exposure limits defined by the American National Standards Institute [40]. The ultrasound detector of the system consists of 256 piezoelectric transducers with a central frequency of 4 MHz, which are equidistantly placed on a circular arc with a radius of 6 cm and an angular coverage of 145°. Ultrasound signals are recorded with a sampling frequency of 40 MHz.

### 5.3 Phantom

We utilized a phantom to test the denoising performance of the trained network on physical measurements, for which the underlying ground truth structure is known. The arrangement of the phantom is shown in the inlay in Fig. 4a. Two plastic tubes with inner diameters of 3.0 mm and 0.86 mm and outer diameters of 3.2 mm and 1.52 mm were filled with ink and imaged cross-sectionally to simulate absorbers of different sizes and at different depths. These tubes were immersed into two layers of agar of slightly different densities mixed with Intralipid (6 ml 20% emulsion / 100 ml water) to mimic light scattering and small variations of the speed of sound distribution in biological tissue. Additionally, a copper plate was integrated into the arrangement as a reference that can be seen in both optoacoustic and ultrasound images.

*5.4 Datasets*

**Table S1. Summary of the structure and the size of the four datasets used in this paper.**

| Name | Description | Train split | Validation split | Test split |
|---|---|---|---|---|
| Dataset-EN | simulated noise-free optoacoustic sinograms | 3000 sinograms | 590 sinograms | 629 sinograms |
| | measurements of pure electrical noise sinograms | 2110 sinograms | 590 sinograms | 629 sinograms |
| Dataset-Ph | Scans of a phantom | - | - | 28 sinograms in 1 multispectral stack |
| Dataset-BC | In-vivo scans of human breast lesions | - | - | 2268 sinograms in 81 multispectral stacks |
| Dataset-GN | simulated noise-free optoacoustic sinograms | 3000 sinograms | 630 sinograms | 700 sinograms |
| | simulated white Gaussian noise sinograms | sinograms with standard deviation uniformly sampled from (0, 0.5] | 630 sinograms with equidistant standard deviations in the interval (0, 0.5] | 21 x 700 sinograms of white Gaussian noise with respective fixed standard deviation $\in \{0, 0.1, 0.2, \ldots, 2\}$. |

Due to small changes in the internal configuration by the manufacturer and changing the location of the imaging device, the Gaussian thermal noise in the breast scans (Dataset-BC) is slightly lower than in the scans of water and the phantom (Dataset-EN and Dataset-Ph). Therefore, experiments on Dataset-BC also challenge the ability of our trained network to accommodate differences in noise characteristics between training and application datasets.

*5.5 Data pre-processing*

We band-pass filtered all recorded signals from 500 kHz – 10 MHz to remove signals outside the bandwidth of the transducers and reduce low frequency responses that would otherwise dominate the contrast in reconstructed optoacoustic images. Additionally, all signals were slightly cropped in the time domain to remove filtering artifacts at the signal boundaries and to make the number of signal samples divisible by 16, as required by the chosen neural network architecture, leading to 1808 time samples for each of the 256 detectors per scan.

*5.6 Signal-to-noise ratio*

For Dataset-EN and Dataset-GN, we used the signal-to-noise ratio (SNR), i.e. the ratio of signal power and noise power, to quantify the noise levels in the signals before and after denoising. We calculated the power $P(s)$ of a whole sinogram $s[d,t]$, $d \in [1,2,\ldots,N_{\text{transducers}}]$, $t \in [1, 2, \ldots, N_{\text{samples}}]$ as

$$P(s) := \frac{1}{N_{\text{transducers}} N_{samples}} \sum_{d=1}^{N_{\text{transducers}}} \sum_{t=1}^{N_{\text{samples}}} s[d,t]^2. \quad (1)$$

Based on equation 1, the SNR of a sinogram $s$ with ground truth noise $s_{noise}$ and inferred noise $s'_{noise}$ is defined as

$$SNR := 10 \log_{10}\left(\frac{P(s - s_{noise})}{P(s_{noise} - s'_{noise})}\right) dB. \quad (2)$$

Setting $s'_{noise} = 0$ yields the SNR before denoising; setting $s'_{noise}$ to the output of the trained network yields the SNR after denoising.

Signal-to-noise ratio in in-vivo scans

Since computing the SNR requires direct access to the ground truth noise $s_{noise}$ of a scan, the metric cannot be directly transferred to the in-vivo scans of Dataset-BC. We therefore defined an alternative metric, $SNR_{mean}$, that enabled us to approximate the ground truth electrical noise for the in-vivo scans from Dataset-BC by considering the per-pixel mean sinogram amplitudes across $N_{scans}$ different scans of the dataset, $s^{(1)}, s^{(2)}, \ldots, s^{(N_{scans})}$, $\langle |s| \rangle := \frac{1}{N_{scans}} \sum_{n=1}^{N_{scans}} |s^{(n)}|$.

$$SNR_{mean} := 10 \log_{10} \left( \frac{P_{\langle |s| \rangle - \langle |s_{noise}| \rangle}}{P_{\langle |s_{noise}| \rangle - \langle |s'_{noise}| \rangle}} \right) dB. \tag{3}$$

We approximated the ground truth electrical noise $\langle |s_{noise}| \rangle$ from the first 100 averaged time samples of all scans in Dataset-BC $\langle |s[t']| \rangle, t' \in [1, 2, \ldots, 100]$:

$$\langle |s_{noise}[d, t]| \rangle \approx \frac{1}{100} \sum_{t'=1}^{100} \langle |s[d, t']| \rangle \text{ for } t \in [1, 2, \ldots, N_{\text{samples}}]. \tag{4}$$

Note that equation 4 yields a meaningful approximation of $\langle |s_{noise}[d, t]| \rangle$ for two reasons: First, the 100 signals recorded at the beginning of a scan do not contain optoacoustic responses but mostly electrical noise because they origin from the coupling medium inside the imaging probe. Second, we observed from the electrical noise sinograms in Dataset-EN that for the used test system $\langle |s_{noise}[d, t]| \rangle$ is constant over time. Thus, an estimation of electrical noise based on a subset of time samples of the sinogram (i.e. the first 100 time samples in equation 4) is applicable to all time steps $t \in [1, 2, \ldots, N_{\text{samples}}]$. Furthermore, sinograms from Dataset-BC were cropped to the first 1732 recorded signal samples before evaluating the $SNR_{mean}$, as subsequent signals originate from outside the designated field of view of the scans and contain strong reflections and filtering artifacts.

## 5.7 Model-based optoacoustic image reconstruction

To evaluate the effects of the denoising method visually and quantitatively on optoacoustic images, we reconstructed the initial pressure $p_0$ of all breast scans in Dataset-BC, both with and without denoising the recorded sinograms with the trained neural network, using a model-based reconstruction algorithm [20, 21]. We added two regularization terms to address the two main causes of the ill-posedness of the inverse problem: simple Tikhonov regularization to mitigate limited view noise and Laplacian-based regularization to mitigate sub-resolution noise.

$$p_0 := \arg\min_{p \geq 0} \|Mp - s\|_2^2 + \lambda_1 \|p\|_2^2 + \lambda_2 \|\Delta p\|_2^2. \tag{5}$$

The reconstructed optoacoustic images are of the size 400 x 400 pixels and correspond to FOVs of 3.99 cm x 3.99 cm. We denote the obtained datasets of reconstructed MSOT breast images as $D_{original}$ and $D_{denoised}$.

Contrast resolution

We quantified the effects of the denoising method in the reconstructed images from $D_{original}$ and $D_{denoised}$ by calculating the contrast resolution of blood vessels. For that, we manually segmented blood vessels in the images, as well as background ROIs from the surroundings of all segmented vessels. The segmentations were based on scans at 870 nm, where blood contrast is at a maximum. The background areas were chosen so as not to overlap with regions below and above strong absorbers, which are affected by limited view artifacts. We chose vessels of different sizes and at different depths to obtain a general estimate for the blood contrast resolution in the dataset. Fig. 3f shows the segmented regions for an exemplary scan. Based on these segmentations, the contrast resolution of a scan with mean intensities $I_{vessels}$ and $I_{background}$ in its respective vessel and background ROIs is defined as

$$CR \coloneqq \frac{I_{vessels} - I_{background}}{I_{vessels} + I_{background}}. \quad (6)$$

*5.8 Blind spectral unmixing via non-negative matrix factorization*

To evaluate the effects of denoising on the spectral contrast of MSOT, we applied blind spectral unmixing via non-negative matrix factorization (NMF) [41] to the multispectral optoacoustic images from $D_{original}$ and $D_{denoised}$ and compared the variety and biological interpretability of obtained spectral decompositions.

For each of the two datasets (consisting each of 400 x 400 x 81 =12 960 000 acquired spectra), we obtained a spectral decomposition into 10 non-negative spectral components H (size 10 x 28) and corresponding non-negative unmixing coefficients W (size 12 960 000 x 10).

$$(W, H) \coloneqq \arg \min_{(W,H) \geq 0} \frac{1}{2} \|S - WH\|_F^2 + \lambda_1(\|W\|_1 + \|H\|_1) + \frac{1}{2}\lambda_F(\|W\|_F^2 + \|H\|_F^2), \quad (7)$$

where S (size 12 960 000 x 28) denotes all spectra of the dataset, $\|M\|_F \coloneqq \left(\sum_{i,j} m_{i,j}^2\right)^{\frac{1}{2}}$ denotes the Frobenius norm, $\|M\|_1 \coloneqq \sum_{i,j} |m_{i,j}|$ denotes the entrywise $L^1$-norm of a matrix $M = (m_{i,j})_{i,j}$, and $M \geq 0$ refers to an entrywise inequality. The entrywise $L^1$-regularization was chosen to promote a maximally sparse decomposition of the spectra, guided by the fact that the spectral contrast of biological tissue is dominated by a small number of abundant chromophores. The number of components and regularization parameters $\lambda_1 = 50.1$ and $\lambda_F = 50.1$ were selected via parameter space exploration and meaningfulness of the resulting spectral components, yielding relative errors $\|S - WH\|_F^2 / \|S\|_F^2$ of 16.8% and 8.2% for the original and denoised data, respectively.

*5.9 Relative contributions of selected NMF spectra per depth*

As a further measure for spectral contrast in $D_{original}$ and $D_{denoised}$, we compare the extent to which depth dependent variations of absorption due to fluence attenuation are captured in the respective spectral decompositions. We first selected the spectral NMF components of both datasets that mainly correlate to the absorption spectrum of hemoglobin, a chromophore that can assumed to be present at all depths. Second, we calculated the average coefficients of these components at depth levels from 0 to 35 mm in steps of 0.1 mm and plotted the progression of these averages as a function of depth. To reduce small fluctuations in the depth plots, we additionally applied a moving average of size +/- 0.8 mm.